\documentclass[a4paper,%
  12pt,%
  abstracton,%
  footexclude,%
  normalheadings,%
  pointednumbers,%
  halfparskip,%
]{scrartcl}

\usepackage{epsfig}
\newcommand{\UPLB}{University of the Philippines Los Ba\~{n}os}

\begin{document}

\title{\large Inferences in a Virtual  Community:\\Demography, User Preferences, and Network Topology}
\author{
   \large Jaderick P. Pabico\\
   \large{Institute of Computer Science}\\
   \large{\UPLB}\\
}
\date{}
\maketitle

\begin{abstract}
This paper presents a computational procedure for extracting demography data, mining patterns of human preferences, and measuring the topology of a virtual network. The network was created from the personal and relationships data of an online Internet-based community, where persons are considered nodes in the network, and relationships between persons are considered edges. A community of Friendster users whose listed hometown is Los Ba\~nos, Laguna was used as a test bed for the methodology. The method was able to provide the following demographic, preferential, and topological results about the test bed:
\begin{enumerate}
\item There are more female users (52.34\%) than male (47.66\%);
\item Ages 15--25 of both genders compose 68\% of the users, with ages 26--40 following at 28\%, ages 41--85 at 4\%, and senior citizens (64--85 years old) at 1\%;
\item Homophily (i.e., birds-of-a-feather adage) is observed in the preferences of users with respect to age levels, such that they are strongly biased towards being friends with users of a similar age; 
\item There is heterophily in gender preference such that friendship among users of the opposite gender occurs more often.
\item The friendship network is well-connected and robust to node removal, such that users can still reach other users through another friend's circle of friends, even if another user leaves the network;
\item It exhibits a small-world characteristic with an average path length of 4.5 (maximum=12) among connected users, shorter than the well-known {\em six degrees of separation}~\cite{travers69}; And
\item The network exhibits a scale-free characteristics with heavily-tailed power-law distribution (with the power $\lambda = -1.02$ and $R^2 = 0.84$) suggesting the presence of many users acting as the network hubs.
\end{enumerate}
The methodology was successful in providing important data from a virtual community which can be used by several researchers in the fields of statistics, mathematics, physics, social sciences, and computer science.
\end{abstract}


\section{Introduction}\label{intro}
In the beginning of the 21st century, the pervasive nature of the Internet has made its way into the lives of the connected humans on the planet through various web services called social networking. Examples of social networking sites are Friendster\footnote{\tt http://www.friendster.com}, MySpace\footnote{\tt http://www.myspace.com}, and Facebook\footnote{\tt http://www.facebook.com}. These web services allow users to publish personal information such as age, gender, relationship status, geographic location, and to mark other users as friends. Thus, these social networking sites provide unprecedented opportunities to capture and analyze the demography, user preference, and the topology of a community on a large scale without resorting to the traditional procedure of surveying a population sample.


In recent years, several efforts have been done to capture and analyze the social structure of virtual communities. Two very recent works that are worth mentioning here are those by Zinoviev~\cite{zinoviev08} and by Leskovec and Horvitz~\cite{leskovec08}. Zinoviev attempted to map the topology and geometry of a Russian social network called {\em Moi Krug}\footnote{{\em Moi Krug} literally means {\em My Circle}} by utilizing several graph theoretic metrics such as node degree distribution and path lengths. In his effort to define the community structure and the boundaries and inner areas of a clich\`e, he introduced the concept of the dense core and the local density. He also used these concepts to  identify the socially popular and marginal users~\cite{zinoviev08}. Leskovec and Horvitz~\cite{leskovec08}, on the other hand, examined and analyzed the characteristics and patterns that emerge from the collective dynamics of large number of people participating in a high-level communication activities using the Microsoft instant-messaging system. From their data, they constructed a communication network with 180 million nodes and 1.3 billion undirected edges. After analysis of the network, they found out that the average degree of separation between nodes agree with the six degrees of separation adage~\cite{leskovec08}.

In this effort, a computerized methodology to extract the demography, user preferences, and network topology of a virtual community was designed, implemented, and tested using as test bed the virtual community of Friendster users from a local municipality. The methodology is centered on a customized web robot designed to crawl and extract the demographic data and other statistics of community of users. The method includes creating an undirected network to mathematically map (and visually graph) the relationship data using nodes and edges to represent the users and their relationships, respectively. From this network, several graph metrics were computed such as the average path length and degree distribution.

Presented in this paper is a methodology for extracting the demographic and preferential patterns of users in a virtual community. Using the 7,172 Friendster users from Los Ba\~nos, Laguna as test bed for the method, the resulting network was analyzed based on centrality metrics. Section~\ref{friendster} briefly describes the test bed social networking site and the kind of data that can be extracted from its database. Section~\ref{extract} describes the web robot employed to extract the demography and preferential data of the selected virtual users, while Section~\ref{results} discusses the results of the extraction, as well as the network analysis performed. Lastly, Section~\ref{summary} summarizes the findings of this study.

\section{Friendster: The Test Bed Social Networking Site}\label{friendster}

Friendster is a web-based social networking service founded in 2002 by Jonathan Abrams~\cite{inoue03}. Based on the {\em Circle of Friends} and {\em Web of Friends} techniques for connecting humans in a virtual environment~\cite{rosen07}, the Friendster network demonstrate the small world phenomenon~\cite{steele98}. Currently, the Friendster database has an estimated 70 Million records corresponding to human users worldwide~\cite{about-friendster}.

The Friendster system uses a method and a device to calculate, display, and perform actions on relationships on the social network. The method, termed the {\em Web of Friends}, combines the methodologies in the {\em Circle of Friends} and the {\em Web of Contacts} to collect descriptive information about their users. The method also allows the Friendster users to tag other users as their friends. Using the method, the descriptive information, as well as the friendship data, are integrated and processed to reveal a series of social relationships connecting any two users in the network. Thus, a specific user can determine the optimal relationship path (i.e., a contact pathway) to reach desired individuals. Moreover, a communications tool was added to allow users to be introduced (or introduce themselves) and initiate direct communication with others~\cite{patent-friendster}.
\section{Extracting Data from a Virtual Community}\label{extract}

\subsection{The Web Robot}

A web robot was created using Perl scripts and a combination of the Linux command line utility programs {\tt grep}~\cite{grep} and {\tt wget}~\cite{niksic}. A Friendster account~$V$ was first created because only currently logged-in users can view the profiles of another Friendster user. Here, $V$~is actually an account of a human Friendster user and is not just a dummy account. Friendster has already purged their database of {\em Pretendsters}, {\em Fakesters}, and {\em Fraudsters}~\cite{terdiman04}, so it is guaranteed that the accounts that the web robot is crawling and extracting data from are that of real and living humans'. The web robot uses the cookie file of the web browser being used by the currently logged-in~$V$. Thus, in the point of view of the Friendster web server, the web robot is nothing but $V$~himself, navigating through the network of friends of Friendster users.

\subsection{Los Ba\~nos Friendster Users}

\begin{figure}[htb]
\centering
\epsfig{file=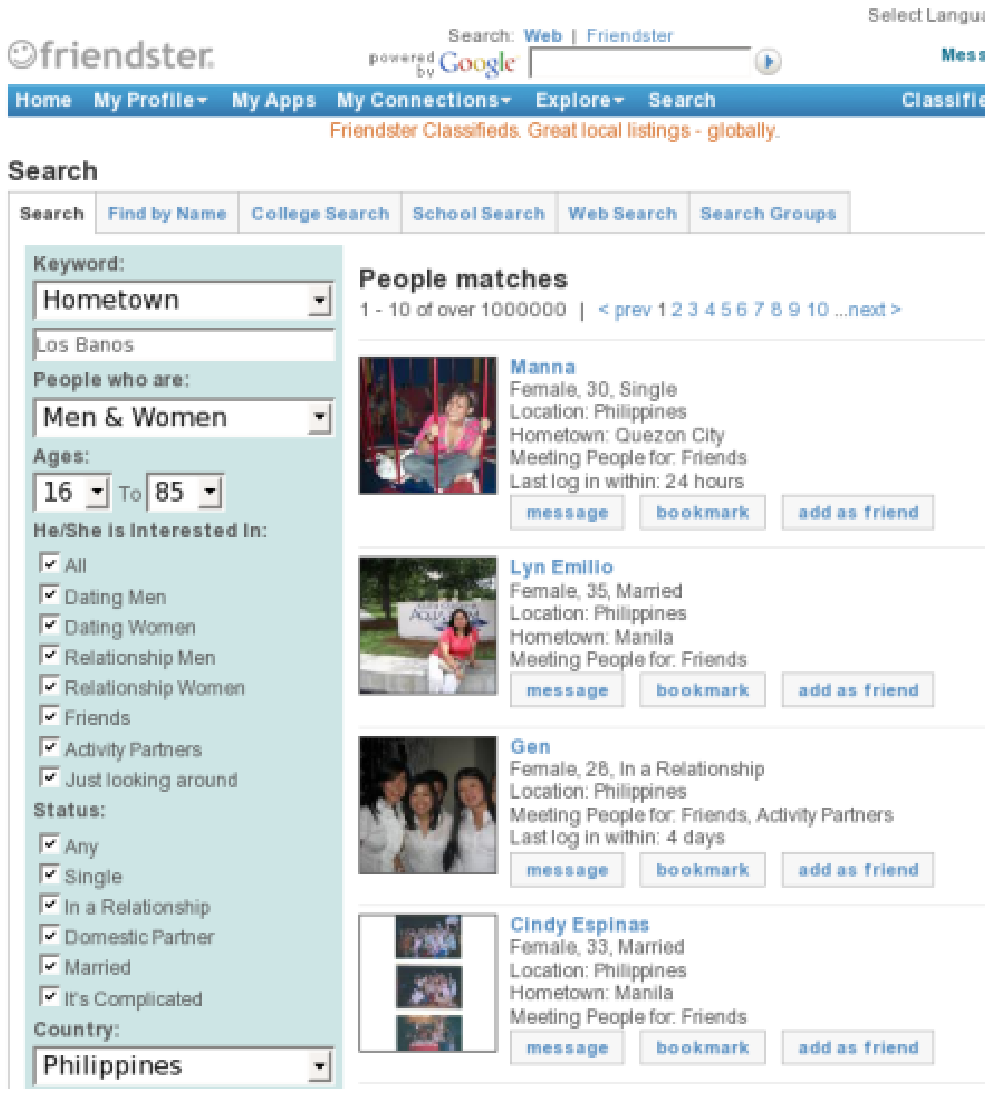, width=3in,height=3.24in}
\caption{A snapshot of the Friendster search tool showing all the available search parameters.}
\label{search}
\end{figure}

The Friendster's search tool was utilized to extract from the Friendster database the accounts of those users whose listed hometown is Los Ba\~nos, Laguna. The search parameter used were those that will list all genders, the widest age range, and all toggle parameters set to on. These toggle parameters are those that refer to friendship preferences and relationship status (Figure~\ref{search}). The Friendster web tool outputs an array $L=\{l_0, l_1, \dots, l_{p-1}\}$ of $p$~pages to list $N$~unique accounts. The pages $l_0, l_1, \dots, l_{p-2}$, each lists 10 unique accounts, while the last page~$l_{p-1}$ lists $p$~modulo $N$~accounts. The web robot started crawling $l_0$~and used its URL to crawl through the succeeding pages~$l_i$, $\forall i=1,\dots,p-1$, by changing only a parameter in the URL. For each page~$l_i$, the web robot automatically extracted the account number, user name, age, gender, and relationship status of each user. These information are then stored in a database table~$\Delta_u$ such that at the end of the crawl, $\Delta_u$~had $N$~unique records corresponding to unique Friendster accounts.

During the web robot crawl, each user's list of friends were also crawled and their respective data extracted and stored in $\Delta_u$. A user's friends list, however, was stored in a separate database table $\Delta_f$, taking note only of the user's account number and the friends' respective account numbers. The tables $\Delta_u$ and $\Delta_f$ are in one--to--many relationship on the user's account number as foreign key.

\subsection{Extracting Demography Data}

The following demographic data were extracted from~$\Delta_u$:
\begin{enumerate}
\item Number~$N_g$ and percentage~$P_g$ of users by gender~$\delta_g$;
\item Number~$N_a$ and percentage~$P_a$ of users by age group~$\delta_a$;
\item Number~$N_r$ and percentage~$P_r$ of users by relationship status~$\delta_r$;
\item Number~$N_{g\times a}$ and percentage~$P_{g\times a}$ of users by~$\delta_g$ and~$\delta_a$;
\item Number~$N_{g\times r}$ and percentage~$P_{g\times r}$ of users by~$\delta_g$ and~$\delta_r$;
\item Number~$N_{a\times r}$ and percentage~$P_{a\times r}$ of users by~$\delta_a$ and~$\delta_r$; and
\item Number~$N_{g\times a\times r}$ and percentage~$P_{g\times a\times r}$ of users by~$\delta_g$, $\delta_a$ and~$\delta_r$.
\end{enumerate}
Here, it is easy to note that the percentage statistics can be derived using the frequency statistics via the following equations:
\begin{eqnarray}
  P_g &=& \frac{N_g}{N}\times 100\%;\nonumber\\
  P_a &=& \frac{N_a}{N}\times 100\%;\nonumber\\
  P_r &=& \frac{N_r}{N}\times 100\%;\nonumber\\
  P_{g\times a} &=& \frac{N_{g\times a}}{N}\times 100\%;\nonumber\\
  P_{g\times r} &=& \frac{N_{g\times r}}{N}\times 100\%;\nonumber\\
  P_{a\times r} &=& \frac{N_{a\times r}}{N}\times 100\%; {\rm and}\nonumber\\
  P_{g\times a\times r} &=& \frac{N_{g\times a\times r}}{N}\times 100\%.\nonumber
\end{eqnarray}
The basic frequency statistics are described as follows:
\begin{enumerate}
  \item $N_g$ is either $N_{\rm male}$ or $N_{\rm female}$. Note that $N_{\rm male} + N_{\rm female} = N$.
  \item $N_a$ is one of $N_{\rm 8--25}$, $N_{\rm 26--40}$, $N_{\rm 41--64}$ or $N_{\rm 65--80}$. Note here that $N_{\rm 8--25}+N_{\rm 26--40}+N_{\rm 41--64}+N_{\rm 65--80}=N$.
  \item $N_r$ is any of $N_{\rm single}$, $N_{\rm married}$, $N_{\rm IAR}$, or $N_{\rm unk}$, where IAR and unk mean {\em in a relationship} and {\em unknown} or disclosed relationship status, respectively. Again, note that $N_{\rm single}+N_{\rm married}+N_{\rm IAR}+N_{\rm unk}=N$.
\end{enumerate}
The compound frequency statistics were just the frequency of users of a given combination of attribute values. For example, the number of single, male users in the age group 8--25 years old can be computed as $N_{\rm male \times 8--25 \times single} = |\delta_{\rm male} \bigcap \delta_{\rm 8--25} \bigcap \delta_{\rm single}|$.
\subsection{Extracting Patterns of Preferences}

The following preferential patterns were extracted from the tables~$\Delta_u$ and~$\Delta_f$:
\begin{enumerate}
\item Preference with respect to gender;
\item Preference with respect to age; and 
\item Preference with respect to relationship status.
\end{enumerate}
To analyze if there are patterns in gender preferences, the frequency of male--male, female--female, and male--female relationships were extracted using an SQL query on the inner join of~$\Delta_u$ and~$\Delta_f$. Similarly, the patterns of age group preferences, as well as that of the relationship status, were extracted using an SQL query on the same inner joins, but on different respective table attributes.

\subsection{Creating and Analyzing the Friendship Network}
The friendship network was created using the data in~$\Delta_f$. Taking each user as node, and the relationship between users as edges, a $N \times N$ adjacency matrix~$R$ was created. The element $r_{i,j}\in R$ takes the value 1 if a relationship record between users~$i$ and~$j$ exists in~$\Delta_f$, otherwise $r_{i,j}=0$.

From~$R$, the following network metrics were computed:
\begin{enumerate}
\item The minimum, average, and maximum number of edges connected to any arbitrary node;
\item The minimum, average and maximum path lengths between two arbitrary nodes; and
\item The degree distribution of the number of edges an arbitrary node has.
\end{enumerate}

\section{Results and Discussion}\label{results}

\subsection{Demography}
\begin{figure}[htb]
\centering
\epsfig{file=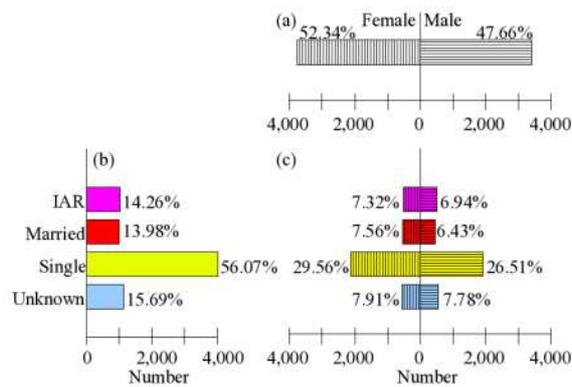, width=3in,height=2.03in}
\caption{(Colored on digital copy) Number of users and percentage (a)~by gender, (b)~by relationship status, and (c)~by gender and relationship status.}
\label{fig1}
\end{figure}

Figure~\ref{fig1} shows the number (and percentage) of users by (a)~gender,  (b)~relationship status, and (c)~relationship status and gender. In Figure~\ref{fig1}a Female users outnumber male users by 4.78\% (336 users). In Figure~\ref{fig1}b, single users dominate the network at 56.07\%. Users who are in IAR, married, or with undisclosed relationship status compose 14.26\%, 13.98\%, and 15.69\%, respectively. In Figure~\ref{fig1}c, female users who are IAR, married, single, and even those with undisclosed relationship status outnumber males by 0.38\%, 1.13\%, 3.05\%, and 0.13\%, respectively.

\begin{figure}[htb]
\centering
\epsfig{file=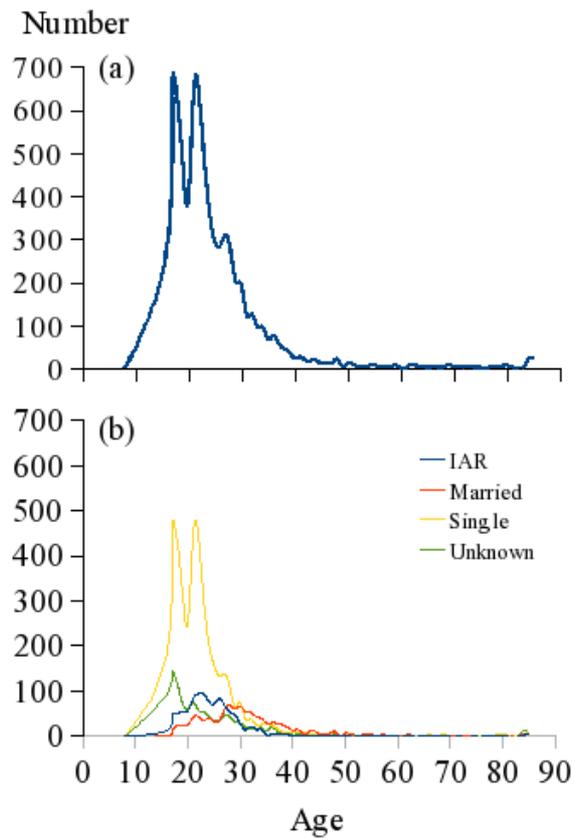, width=3in,height=4.47in}
\caption{(Colored on digital copy) Number of users (a)~by age, and (b)~by age and relationship status.}
\label{fig2}
\end{figure}

\begin{figure}[htb]
\centering
\epsfig{file=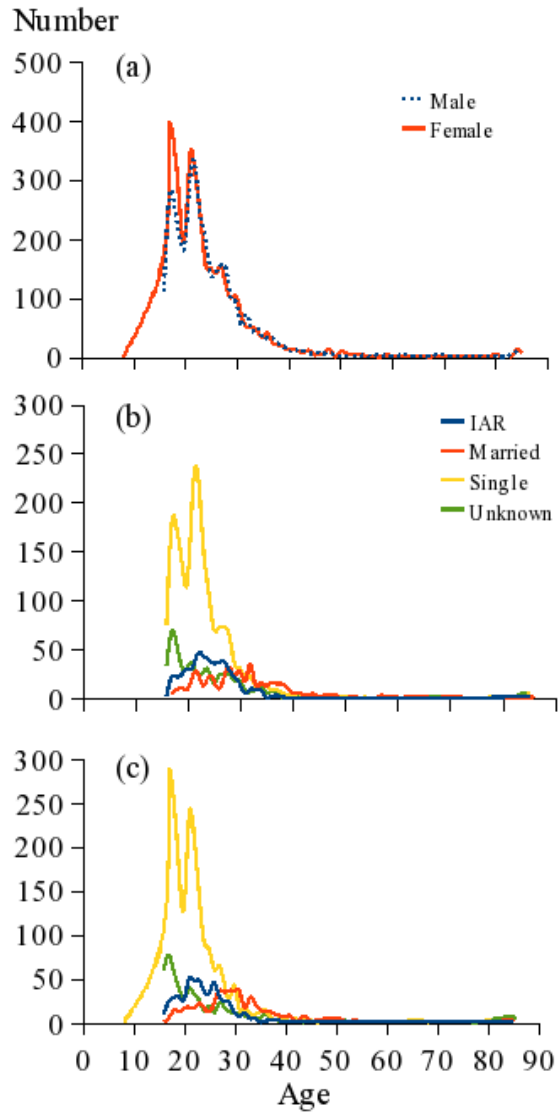, width=3in,height=6in}
\caption{(Colored on digital copy) Number of (a)~female and male users by age, (b)~female users by age and relationship status, and (c)~male users by age and relationship status.}
\label{fig3}
\end{figure}

Figure~\ref{fig2} shows the number of users (a)~by age and (b)~by age and relationship status. The trend shows that the dominant users are in the late teens and mid-twenties as shown by the two, equally-high peaks. This pattern is consistent with the pattern of the single users. Among the married users, late 20's and early 30's are the dominant age. Among the users who are IAR, the dominant age is early 20's. Figure~\ref{fig3} shows the number of users (a)~by age and gender, and (b-c)~by age, gender, and relationship status. The trend among female users generally follows that of the single's, but there are more users among the late teens than that of the early 20's. The trend among male users is also somewhat similar to that of the single's, but there are more users among the early 20's than that of the late teens.

\subsection{Patterns of Friendship Preferences}

\begin{figure}[ht]
\centering
\epsfig{file=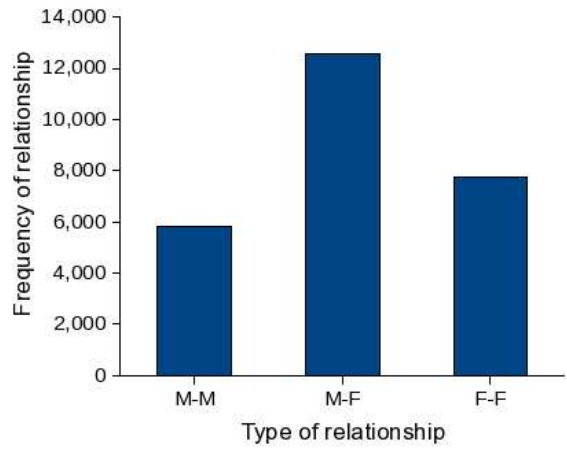, width=3in,height=2.4in}
\caption{Frequency of male-male, male-female, and female-female relationships (where M=male and F=female).}
\label{fig4}
\end{figure}

The pattern of friendship preferences with respect to gender was extracted from the network created. The frequency of friendships between two male users, between two female users, and between a male and a female user were counted. Figure~\ref{fig4} summarizes the results of the frequency analysis. The figure shows that relationships with opposite gender occur more often than with the same gender. Thus, it can be deduced that users prefer to be in a relationship with opposite gender, suggesting heterophily in gender preferences.

\begin{figure}[ht]
\centering
\epsfig{file=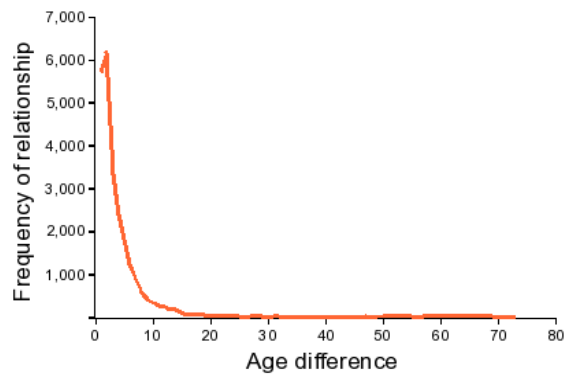, width=3in,height=2in}
\caption{Frequency of relationships between users with respect to age differences.}
\label{fig5}
\end{figure}

Figure~\ref{fig5} shows the frequency of relationships between users whose age difference ranges from 0 to 75 years. The pattern shows that relationships with users whose age is at most 10 years occur more often than whose age difference is greater than 10 years. Thus, it can be deduced from the pattern that users select to be friends with users of the same age, suggesting homophily in age level preference.

\begin{figure}[ht]
\centering
\epsfig{file=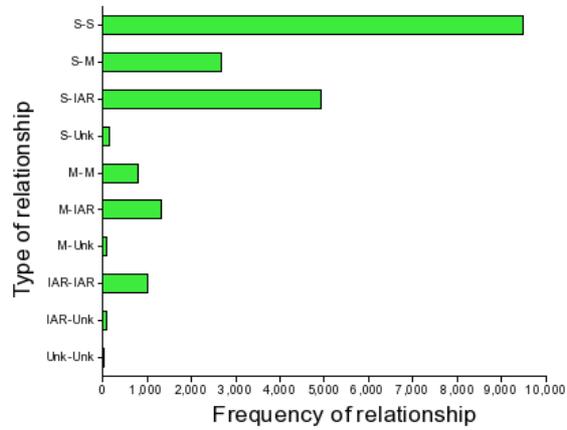, width=3in,height=2.3in}
\caption{Frequency of relationships with respect to relationship status (where S=single, M=married, IAR=in a relationship, and Unk=unknown).}
\label{fig6}
\end{figure}

Figure~\ref{fig6} shows the frequency of relationships between users with respect to relationship status. The pattern shows that half of the singles prefer to be friends with singles (i.e, homophily) while half prefer to be friends with people who are already IAR (i.e., heterophily); IAR users prefer to be friends with those who are already IAR.

\subsection{Centrality and Topology of the Social Network}

\begin{figure}[ht]
\centering
\epsfig{file=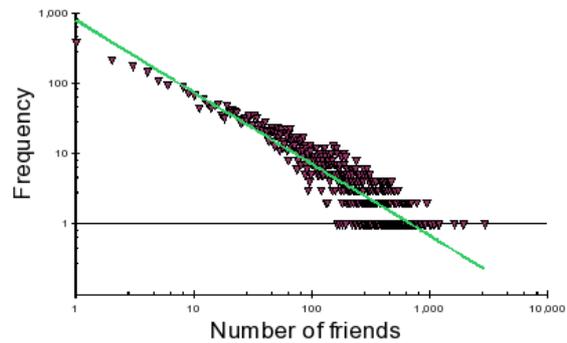, width=3in,height=1.88in}
\caption{Log-log plot of the number of friends $\times$ frequency obeys the power law distribution. The straight line is the power law line fit whose slope was found to be $\lambda=-1.02$ with $R^2=0.84$.}
\label{fig7}
\end{figure}

The results of path analysis show that the network has a maximum and an average path length of 12 and 4.5, respectively. This suggests that one user of the network can reach another user through a friend of a friend via an average of 4.5 persons, and that the person is guaranteed to be reached via a minimum of 12 persons. Figure~\ref{fig7} shows the distribution of the number of friends (number of friends $\times$ frequency) in the log-log scale. The distribution obeys the power law distribution, thus the network is considered scale-free. The presence of heavy tail in the distribution suggests that many users are acting as the network hubs. This implies that information sources must target these hubs because they have a wide sphere of influence as they are considered opinion leaders. Information (or gossip, or epidemic) spread faster into the network if they originate from these hubs.

\section{Summary and Conclusion}\label{summary}
This paper presents a computational methodology for extracting demography data, mining patterns of human preferences, and measuring the topology of a virtual network. Using the Friendster network of Los Ba\~nos residents as a test-bed for the methodology, the following facts were extracted:
\begin{enumerate}
\item There are more female community users (52.34\%) than male (47.66\%);
\item Ages 15--25 of both genders compose 68\% of the users, with ages 26--40 following at 28\%, ages 41--85 at 4\%, and senior citizens (64--85 years old) at 1\%;
\item Homophily (i.e., birds-of-a-feather adage) is observed in the preferences of users with respect to age levels, such that they are strongly biased towards being friends with users of a similar age; 
\item There is heterophily in gender preference such that friendship among users of the opposite gender occurs more often.
\item The friendship network is well-connected and robust to node removal, such that users can still reach other users through another friend's circle of friends, even if another user leaves the network;
\item It exhibits a small-world characteristic with an average path length of 4.5 (maximum=12) among connected users, shorter than the well-known {\em six degrees of separation}~\cite{travers69}; And
\item The network exhibits a scale-free characteristics with heavily-tailed power-law distribution (with the power $\lambda = -1.02$ and $R^2 = 0.84$) suggesting the presence of many users acting as the network hubs.
\end{enumerate}
The methodology was successful in providing important data from the test-bed virtual community. This method can be used by several researchers in the fields of statistics, mathematics, physics, social sciences, and computer science.

\section{Acknowledgments}

This work was supported by (CHED's National Center of Excellence in Information Technology Education) Institute of Computer Science, College of Arts and Sciences, \UPLB\ through UPLBGF \#2326103 and UPLBFI \#2004987. The Los Ba\~nos Friendster network was extracted using the 256-node adhoc computing grid of the Institute of Computer Science.

\bibliography{methodology}
\bibliographystyle{abbrv}

\end{document}